# Right-sizing compute resource allocations for bioinformatics tools with Total Perspective Vortex

Nuwan Goonasekera[1], Catherine Bromhead[1], Simon Gladman[1], Nate Coraor[2], Bjorn Gruning[3], Enis Afgan[4]

[1] Melbourne Bioinformatics, University of Melbourne, Melbourne, Australia
[2] Pennsylvania State University, State College, PA, USA
[3] University of Freiburg, Freiburg, Germany
[4] Johns Hopkins University, Baltimore, MD, USA

## Abstract

In biomedical research, computational methods have become indispensable and their use is increasing, making the efficient allocation of computing resources paramount. Practitioners routinely allocate resources far in excess of what is required for batch processing jobs, leading to not just inflated wait times and costs, but also unnecessary carbon emissions. This is not without reason however, as accurately determining resource needs is complex, affected by the nature of tools, data size, and analysis parameters, especially on popular servers that handle numerous jobs. The Galaxy platform, a web-based hub for biomedical analysis used globally by scientists, exemplifies this challenge. Serving nearly half a million registered users and managing around 2 million monthly jobs, Galaxy's growth outpaces the resources at its disposal. This is necessitating smarter resource utilization. To address this, we have developed a tool named Total Perspective Vortex (TPV) - a software package that right-sizes resource allocations for each job. TPV is able to dynamically set resource requirements for individual jobs and perform meta-scheduling across heterogeneous resources. It also includes a first-ever community-curated database of default resource requirements for nearly 1,000 popular bioinformatics tools. Deployments in Galaxy Australia and Europe demonstrate its effectiveness with meta-scheduling user jobs and an improved experience for systems administrators managing Galaxy servers.

## Introduction

The ongoing growth trend in biological data sciences is putting pressure on compute resource availability. The total number of researchers relying on analytics is growing; the number of active users is growing; the number of processing jobs is growing (Fig. 1). For public service providers, such as Galaxy [1], this growth is causing longer queue wait times for users. For commercial cloud computing users, this growth is causing higher infrastructure bills. This growth in analytical methods also carries a substantial carbon footprint [2]. While analytical methods are a key tool in modern biomedical research and these side effects are a part of the process, there are opportunities to improve, and ideally optimize, use of resources by right-sizing them for the job at hand. By right-sizing job resource allocations, service providers maximize the available infrastructure capacity and users get the benefit of balanced time and cost tradeoff.

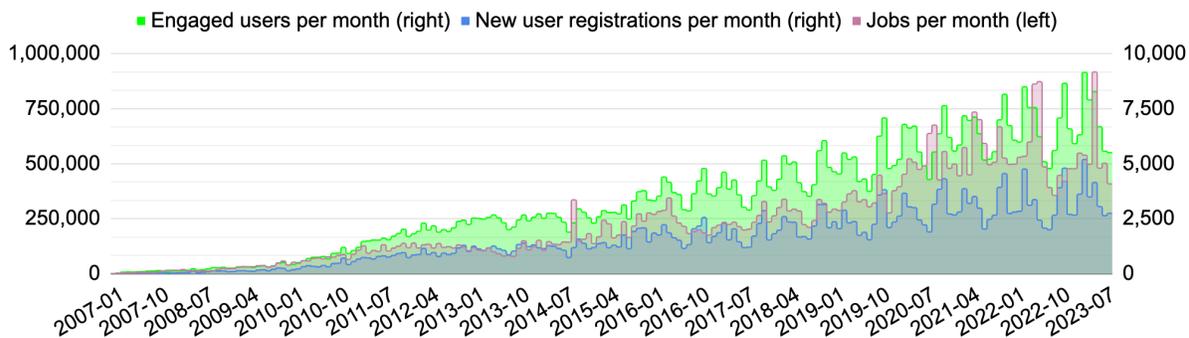

**Figure 1**. Growth across the number of new users, number of active users, and the number of processed jobs on one of the most popular, free bioinformatics servers in the world, *usegalaxy.org*.

Right-sizing means appropriately allocating resources for the job at hand and selecting a suitable target machine. Each analytical job on a batch system requires a resource allocation to be specified before job submission. This typically means the number of CPUs and amount of memory but can also include additional parameters, such as number of GPUs, processor architecture, or use of a specific queue. These parameters are often nuanced to set correctly because they depend on the tool, the size and type of input data, the target system, and even job parameters. As a result, most users, including professional systems administrators, tend to guess these values. And because requesting too few resources can lead to a job running excessively long or, worse, the job terminating before completion, users generally overestimate the value of these properties. Collectively, this leads to a resource overallocation and ultimately waste of resources [3].

In this paper, we describe a software package for interactively right-sizing job resource requests and efficiently assigning jobs to available resources. Written as a plugin module for the widely popular Galaxy application - the Total Perspective Vortex, or TPV - is able to dynamically set resource requests and route jobs to appropriate compute resources based on a set of configurable rules. The rules include tools, users, roles, and destinations that can be interactively reasoned about through a flexible and intuitive tagging system and mapped onto the available resource capacity. TPV also supports writing custom Python-based rules for making more complex decisions. The package is accompanied with a public database of default resource requirements for nearly a thousand bioinformatics tools. This relieves system administrators from needing to learn and make resource decisions for those tools on their own. In our experience, adoption of TPV on large, public, free Galaxy services has led to a substantial Quality-of-Life improvement for systems administrators and a 5% reduction in average job processing. In this paper, we highlight the benefits of TPV across user roles in the bioinformatics community, describe its functionality, and present the shared tool resource database as a new resource for systems administrators in life sciences. TPV is open source and easily configurable for any Galaxy server. The shared tool resource database can be utilized independent of TPV and can hence be integrated with other workflow management systems.

## Value of TPV to the bioinformatics community

In this section we take a look at three examples of where TPV has been adopted and how it has yielded value for end users, systems administrators, and national infrastructure providers. These examples further motivate the need for a solution such as TPV. Technical and implementation details of the features enabling these benefits are described in subsequent sections.

### Improving throughput at usegalaxy.*

Galaxy (https://galaxyproject.org) is a mature, open source workbench for scientific computing, capable of managing and executing complex workflows on a variety of compute infrastructures. One of the unique elements of the Galaxy project is that it offers free public access to substantial compute and storage resources via its *usegalaxy.\** federation. usegalaxy.* is a set of managed Galaxy services distributed around the world. Three main instances operate: one in Australia (usegalaxy.org.au), one in EU (usegalaxy.eu), and one in the US (usegalaxy.org). Collectively, these services have nearly 450,000 registered users, offer access to several thousand bioinformatics tools, and process about 2 million jobs each month.

Behind the scenes, usegalaxy.* services leverage swaths of geographically distributed, heterogeneous compute resources to process users' jobs. These resources come from various national infrastructure initiatives, such as NeCTAR in Australia [4], Elixir in Europe [5], and ACCESS in the US [6]. The resources are provided through a set of allocations that the local groups apply for and manage. Consequently, there is limited compute capacity available and it is important to utilize these allocations to the best extent possible when running user jobs.

In the past, resource allocation for user jobs had been done through a combination of custom solutions and administrator-defined values. This included defining values for the Galaxy application's Dynamic Tool Destinations implementation [7], a Python script called the Job Router that was developed specifically for usegalaxy.org [8], and a more generic job sorting implementation known as Sorting Hat [9]. These solutions were site-specific and had a number of limitations related to expressivity (e.g., lacking support for job resubmission or resource rejection). They also lacked a global, live view of all the available resources (more on this in the meta-scheduling section below).

To address these issues, in mid-2022, the usegalaxy.* federation started adopting TPV as the default scheduler for all the user jobs. While this change has technically been invisible to end users, the runtime of user jobs has decreased allowing more jobs to be processed using the same allocation (see Results section). The benefits and implications of this are having an impact on the end user experience across the federation.

### Saving time for systems administrators with a tool resource database

Driven by new data types, better understanding of biological processes, and novel algorithms, the bioinformatics domain relies on a constant influx of new tools. For example, BioConda is a popular packaging and distribution framework for bioinformatics tools [10]. It consists of >12,000 tools, with many being updated or new ones added every day (https://github.com/bioconda/bioconda-recipes). The Galaxy ToolShed, a bioinformatics "app store" for Galaxy, is closely tied to BioConda and it contains >9,500 tools. Before any of these tools can be used by end users, they must be installed and configured for use on a given Galaxy instance. A Galaxy administrator tends to manage dozens to hundreds of these tools and is tasked with configuring job submission properties for each one. With the diversity and number of tools, this is a time consuming, imperfect task. And with hundreds of local Galaxy installations all over the world, it is highly duplicative across admins because they each have to set those values for a highly overlapping set of installed tools.

As part of TPV, we have formed a shared resource requirements database that captures sensible default resource values for about a thousand most commonly used bioinformatics tools. Systems administrators can simply point their Galaxy configuration at this database and TPV will automatically apply relevant values for jobs at submission. TPV supports site-specific overrides of the values as well as use of scaling factors to restrict maximum size of resources without having to override each tool independently. The default resource values were derived from historic job data from the usegalaxy.* federation and are being constantly reevaluated. As the global values adjust, Galaxy instances automatically start using new values without requiring any action on behalf of the admin. The database is maintained in the open, on GitHub, so anyone is encouraged to share their experiences and contribute to the value definitions.

**First bioinformatics-focused meta-scheduler**

The resources behind usegalaxy.* federation tend to be shared, heterogeneous, and geographically distributed. They include use of clusters dedicated to Galaxy, academic and commercial cloud providers, and access to a number of specialized supercomputers. Galaxy hence operates in a highly dynamic environment with a constant flux of available resources and queue wait times. Coupled with the growing volume and diversity of the workload (Fig. 1), it is important to manage the overall system systematically, with a wholesome view of the entire environment.

TPV has hence been implemented with meta-scheduling capabilities. Meta-scheduling considers a list of potential destinations for a job, relying on live resource load and specific job properties. TPV runs those values through a rank function to order the potential job destinations. The rank function considers system status and any scheduling requirements specified for the tool, such as the values for CPUs and memory as well as any additional requirements, such as use of certain container technology or a runtime environment (more on this in the implementation section). In addition to the default ranking function, TPV supports custom rank functions. Custom rank functions allow a Galaxy administrator to consider additional, site-specific data points before a job is scheduled. Custom data points to consider can be anything, with examples including the type of hardware available, the trend of job submissions over the past X minutes, or the geographic proximity of input data, to name a few.

Combined with the tool resource database, this meta-scheduling capability enables a Galaxy administrator to make efficient use of all available resources with minimal manual interventions. With usegalaxy.* servers, TPV has processed several million jobs demonstrating that the ranking function is efficient and scalable. Even for smaller Galaxy installations with less resource heterogeneity, with TPV being aware of user roles and job sizes, it enables fit-for-purpose scheduling decisions to be made in a straightforward and consistent way.

## TPV implementation and functionality

In this section we describe the library implementation details and the evaluation path that TPV exercises for each job scheduling decision. In the process, we showcase how expressive the implementation is and, with a running example, demonstrate its use. Thorough documentation about how to install and configure TPV is available online: https://total-perspective-vortex.readthedocs.io/

### Defining scheduling resources

TPV is driven via YAML configuration files that allow administrators and TPV to reason about the resources in play. Within TPV, these resources are referred to as Entities. An Entity represents any resource or object that can be considered for a scheduling decision. Entities include Tools, Users, Groups, and Destinations. As the name implies, a Tool captures a set of resource requirements that a software tool requires for execution, such as the number of CPUs or amount of memory. User and Group definitions allow any specific requirements to be assigned to a given username or a role, as defined in Galaxy. These definitions allow specific entities access to an exact Destination or an administrator to limit the maximum size of available resources. An example of use is the ability to route workshop participant jobs to a dedicated queue without wait time. Destinations refer to the resources available for jobs to execute on, such as clusters or specific queues. A Destination also defines any resource restrictions, such as maximum number of CPUs available (more on this below). An example of a configuration definition for Tool and Roles entities is shown in Fig. 1.

```yaml
tools:
  toolshed.g2.bx.psu.edu/repos/iuc/hisat2/.*:
    cores: 12
    mem: cores * 4
    gpus: 1
    env:
      _JAVA_OPTIONS: -Xmx{int(mem)}G -Xms1G
roles:
  training.*:
    cores: 5
    mem: 7
```

**Figure 1**. A sample definition of TPV Tools and Roles entries. The definition specifies the number of CPU cores, amount of memory, the number of GPUs, and job environment details for the HISAT2 tool. As shown with the memory and environment definitions, TPV supports interpolation of values used in the definition itself. The Roles definition specifies that any group whose name starts with `training` will have its cores and memory limited to the specified values.

All Entities have a set of common properties. These include a unique entity id, number of cores, amount of memory, environment details, specific parameters, and scheduling tags. These properties capture the resource capabilities or requirements as well as environment configuration and scheduling constraints for that Entity. The Destination entity has additional properties not possessed by other entity types that further aid scheduling. These include the maximum size of a job accepted at that destination across available resource components (i.e.,

CPU, memory, GPU). In addition to the explicit properties, most entities can also have `rules` property defined. Rules provide a means by which to conditionally change entity properties. They support conditional logic and allow just-in-time scheduling decisions based on a job's requirements and resource state. For example, a Tool rule can inspect the size of the job input and, based on the value, set properties such as number of CPUs, memory, or a Destination. Finally, Entities can inherit information from other entities of the same type (e.g., a Tool can inherit from another Tool). A common practice is to define a default base set of properties that all other entities inherit from, allowing the administrator to only define differing properties. This reduces repetition of defined values and eases sweeping updates. An example with these concepts is shown in Fig. 2.

```yaml
global:
  default_inherits: default
tools:
  default:
    cores: 2
    mem: 4
  toolshed.g2.bx.psu.edu/repos/iuc/hisat2/.*:
    cores: 8
    mem: cores * 4
    gpus: 1
    rules:
      - if: input_size <= 10
        cores: 4
        mem: cores * 4
destinations:
  my_slurm_cluster:
    params:
      nativeSpecification: "--nodes=1 --ntasks={cores} --ntasks-per-node={cores} --mem={mem*1024}"
```

**Figure 2**. An example definition of a Tool entity demonstrating use of property inheritance and a conditional rule. All tool definitions inherit from the `default` definition. In this case the `HISAT2` tool overrides definitions for the `cores` and `mem` while also adding a definition for `gpus`. The `params` property is inherited from the `default` definition. The `if` rule updates those values if the total size of input files is less than 10GB.

### Making scheduling decisions

The resource definitions designate all the properties "suppliers" and "consumers" possess. The next step in the scheduling process is using those property values to make scheduling decisions. TPV is designed to capture three aspects of scheduling in its decision-making process:

a. Resource requirements: these are the amounts of cores, memory, and GPUs that are needed to execute a job;
b. Scheduling constraints: these capture any restrictions or limits on resource use, such as a resource being offline or being available only to specific users;

c. Runtime environment configuration: these are destination-specific arguments that are required by the specific site to run jobs. Examples include a native specification for a cluster management system or environment variables.

The scheduling process stacks, or merges, these to yield a singular scheduling decision with all the necessary properties.

The first step in this scheduling process is to determine whether the target compute destination has sufficient capacity to execute a job. The next step, if there are multiple compute targets, is to find the best fitting destination. This is done by maximizing some utility function, such as overall system throughput, job turnaround time, or some other metrics of choice.

In TPV, the scheduling decision is driven by tags. Scheduling tags determine how entities match up against each other, and can be used to indicate preference or aversion to another entity. Each tag has a name and a category. The name is arbitrary but needs to be unique as it is used to match entities. The category can be one of four possible values: *required*, *preferred*, *accepted*, or *rejected*. These categories range from strong preference to another tag (required) to strong aversion (rejected). For example, a tool entity can define a scheduling tag named "high-mem" and set its category as "required". That tool will then only match up with a destination entity that also has a tag named "high-mem", hence requiring it. Similarly, a tool can define a tag named "offline" as "rejected". Any destination marked with an "offline" tag would then no longer be considered for scheduling. *Preferred* and *accepted* tags are used when ranking destinations. Through creative use of tag categories, complex scheduling behavior can be produced. Table 1 capturing the tag compatibility to facilitate these scheduling decisions.

**Table 1**. Compatibility table for the scheduling tags. Two tags must have a ✓ for the two entities to be able to match during the scheduling decision process.

| Tag Type | Require | Prefer | Accept | Reject | Not Tagged |
|---|---|---|---|---|---|
| **Require** | ✓ | ✓ | ✓ | ✗ | ✗ |
| **Prefer** | ✓ | ✓ | ✓ | ✗ | ✓ |
| **Accept** | ✓ | ✓ | ✓ | ✗ | ✓ |
| **Reject** | ✗ | ✗ | ✗ | ✗ | ✓ |
| **Not Tagged** | ✗ | ✓ | ✓ | ✓ | ✓ |

For more complex scheduling rules, TPV also supports Python expressions and code blocks in the `rules` property. The rules code block has context passed to it that includes the total size of job inputs as well as properties of the Role, Tool, and User entities. These can be used to make resourcing and scheduling decisions of arbitrary complexity with the only requirement being that the block must evaluate to a boolean value.

### Job dispatch process

With all the configuration blocks in place, TPV is able to make a scheduling decision. This is referred to as a job dispatch process, and it follows a fixed procedure to determine the most suitable destination for the job. The following are the steps of this process, visualized in Fig. 3:

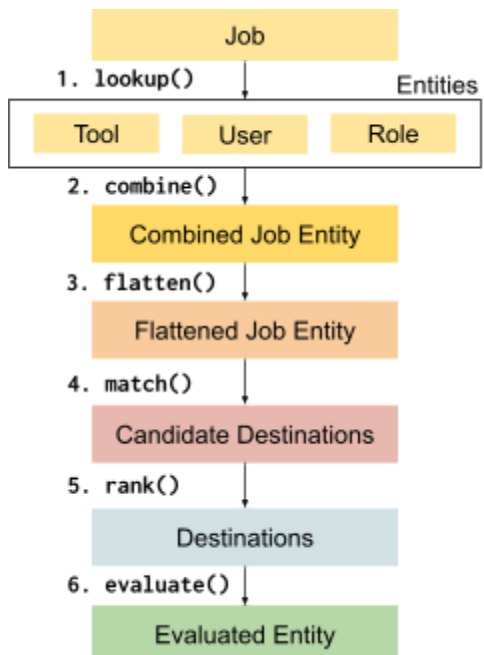

**Figure 3**. Visual representation of the job dispatch process capturing all the evaluation functions and entities.

In order to visualize the operations that entities follow, and the in-memory state as it proceeds through this process, we will use the following running example.

```yaml
tools:
  default:
    env:
      _JAVA_OPTIONS: -Xmx{int(mem)}G -Xms1G
  toolshed.g2.bx.psu.edu/repos/iuc/hisat2/.*:
    cores: 8
    mem: cores * 3
    gpus: 1
    rules:
    - if: input_size <= 0.2
      id: hisat_small_input_rule
      cores: 6
roles:
  training.*:
    cores: 5
    scheduling:
      require:
        - training
```

```yaml
destinations:
  my_slurm_cluster:
    params:
      nativeSpecification: "--nodes=1 --ntasks={cores} --ntasks-per-node={cores} --mem={mem*1024}"
    scheduling:
      accept:
        - training
  my_pulsar_cluster:
    max_accepted_cores: 4
```

**Figure 4**. A sample TPV configuration to motivate the processing steps involved during entity dispatch. We consider a user who dispatches a tool with id: https://toolshed.g2.bx.psu.edu/repos/iuc/hisat2/1.0.0 with roles "training_2023" with an input size of 100MB (0.1 GB).

1. *lookup()* - TPV first looks up all entities in the YAML configuration that are relevant to scheduling the job. For example, the lookup may find that only a Tool and User entities are defined and relevant to the current job, and no Role is relevant. In our example, the entry for entity hisat2 will match, as will the training role. Note how the default inheritance has already been applied to the tool entity.

    ```yaml
    toolshed.g2.bx.psu.edu/repos/iuc/hisat2/1.0.0:
      cores: 12
      mem: cores * 4
      gpus: 1
      env:
        _JAVA_OPTIONS: -Xmx{int(mem)}G -Xms1G
      rules:
        - if: input_size <= 0.2
          id: hisat_small_input_rule
          cores: 6
    training.*:
      cores: 5
      scheduling:
        require:
          - training
    ```

    **Figure 5**. The in-memory state of the matching entities.

2. *combine()* - The relevant entities are merged into a single Job entity, following a predefined order of stacking. The combined entity shares all the respective preferences and rules of its constituent entities. If the same property is defined on both entities, the entity with the higher merge priority will override the other. The priority order is fixed in the following way: Destination > User > Role > Tool. For example, if a Tool specifies a value for `cores`, and a User also specifies `cores`, the User entity `cores` value will take precedence.

Scheduling tags are also merged at this point. For example, if a matching `role` declares a tag named "training", and a matching Tool expresses a requirement for a tag named "high-mem", the combined entity would share both preferences. This can be used to route certain roles or users to specific destinations for example.

However, if the tags are mutually exclusive, scheduling cannot occur. For example, if a Role expresses a *preference* for "training" tag, but the Tool "rejects" tag "training", then the Job cannot be scheduled. If the tags are compatible, then the tag with the stronger claim takes effect. For example, if a Tool *requires* "high-mem" and a User *prefers* "high-mem", then the combined entity will *require* "high-mem".

```yaml
combined:
  cores: 5
  mem: cores * 4
  gpus: 1
  env:
    _JAVA_OPTIONS: -Xmx{int(mem)}G -Xms1G
  rules:
    - if: input_size <= 0.2
      id: hisat_small_input_rule
      cores: 6
  scheduling:
    require:
      - training
```

**Figure 6**. The in-memory state of the combined entities.

3. *flatten()* - Evaluates any conditional clauses on the Job entity, flattening the entity into a single set of expressions.

    ```yaml
    combined:
      cores: 6
      mem: cores * 4
      gpus: 1
      env:
        _JAVA_OPTIONS: -Xmx{int(mem)}G -Xms1G
      scheduling:
        require:
          - training
    ```

    **Figure 7**. The in-memory state of the flattened entities. Note how the conditional clauses are evaluated and merged with the parent entity.

4. *match()* - Matches the flattened Job entity requirements with a suitable Destination. This step ensures that the Destination has sufficient resources (`cores`, `mem`, and `gpus`) to satisfy the Job entity's request. Any Destinations that do not have scheduling tags that are *required* by the Job entity are skipped, and all Destinations that have scheduling tags *rejected* by the Job entity are also skipped. *Preference* and *acceptance* tags are not yet considered at this stage.

```yaml
combined:
  cores: 6
  mem: 24
  gpus: 1
  env:
    _JAVA_OPTIONS: -Xmx{int(mem)}G -Xms1G
  scheduling:
    require:
      - training
```

**Figure 8**. The in-memory state of entities during matching. Note how resource related expressions have been evaluated (in this example, mem), so that a destination with sufficient resources can be found. Non resource-related expressions are not evaluated. During the match, the `my_pulsar_cluster` destination is not a match, as the tool requires 6 cores, and `my_pulsar_cluster` only accepts a maximum of 4 core jobs. The scheduling tags are also matched at this stage, and the only destination that fulfills all criteria is the `my_slurm_cluster` destination.

5. *rank()* - After the matching Destinations are shortlisted, they are ranked using a pluggable `rank` function. The default `rank` function sorts the Destinations by tags that have the highest number of *preferred* tags, with a penalty if *preferred* tags are absent. This default rank function can be overridden if desired, allowing a custom rank function to be defined using Python code. This can be used to implement arbitrary logic for picking the best match from the available candidate destinations based on some utility function. This can be used to implement meta-scheduling capabilities by, for example, querying the current load of multiple resources to determine the least loaded one. Figure 4 showcases such an implementation.

```yaml
tools:
 default:
   cores: 2
   mem: 8
   rank: |
import requests
params = {
  'pretty': 'true',
  'db': 'pulsar-test',
  'q': 'SELECT last("percent_allocated") from "sinfo" group by "host"'
}
try:
  response = requests.get('http://stats.genome.edu.au:8086/query', params=params)
  data = response.json()
  cpu_by_destination = {s['tags']['host']:s['values'][0][1] for s in
```

```python
    data.get('results')[0].get('series', [])}
    # Sort by destination preference, and then by cpu usage
       candidate_destinations.sort(key=lambda d: (-1 * d.score(entity),
   cpu_by_destination.get(d.dest_name)))
       final_destinations = candidate_destinations
    except Exception:
       log.exception("An error occurred while querying Influxdb. Using a weighted random candidate
   destination.")
       final_destinations = helpers.weighted_random_sampling(candidate_destinations)
    final_destinations
```

**Figure 9**. An implementation of a custom `rank` function that queries an Influx database service to obtain current load of available Destinations. Based on that data, the function ranks the Destinations by destination preference and then by cpu usage.

6. *evaluate()* - Once the highest ranked destination is known, the Job entity is once again combined with the Destination so that any Destination configuration and resource constraints can also be taken into account. Any remaining Python expressions are evaluated at this point, creating the final set of resource requirements.

```yaml
combined:
  cores: 6
  mem: 24
  gpus: 1
  env:
    _JAVA_OPTIONS: -Xmx6G -Xms1G
  params:
    nativeSpecification: "--nodes=1 --ntasks=6 --ntasks-per-node=6 --mem=24576"
  scheduling:
    require:
      - training
```

**Figure 10**. The final entity state after being combined with the matching destination. Note how the nativeSpecification param comes from the chosen destination (`my_slurm_cluster`), and how all expressions are evaluated to their final values.

Once the dispatch process completes, TPV returns the destination name and set of resource requirements for the job. Based on the combination of the tool resource requirements, the job context, and the current resource status, the returned values represent the right-sized resource requirement for the given point in time. Galaxy then uses these values to formulate a job request for the target resource and submits the job.

## Shared tool resource database

As evident from the job dispatch process, tool resource requirements play an important role in identifying a suitable destination for a job. As highlighted in the introduction, the field of bioinformatics has a high number of tools that are regularly needed and keeping up with an (accurate) set of tool-specific resource requirements is an ongoing, arduous task. To ease this process for systems administrators tasked with defining these values, we have extracted general tool resource requirements and runtime configuration into a separate shared database and published it on GitHub [11]. This database currently defines entries for nearly a thousand tools. The database enables admins to instantly avail themselves of well-known tool resource requirements by simply including the URL of the shared database file in TPV's configuration. In this section, we describe how the database was populated and how it is maintained. We also describe how TPV is designed to maximize the ease with which the database can be adopted by institutions and deployments with various compute capacity.

**Default database values**

The resource requirements database was initially seeded by pooling together existing knowledge on resource requirements across the 3 main deployments of the usegalaxy.* federation. These resource requirements were collected over a decade and capture the experience of running these instances. While the database mostly captures a fixed set of CPU and memory requirements, a few tools also have resource scaling rules defined based on job parameters (e.g., `ncbi_blastp_wrapper`) and a few more have rules based on job input size (e.g., `hifiasm`). Because it is a globally shared database, it was designed to exclude environment-specific optimizations, such as special optimizations for specific instruction sets or DRM settings specific to a local environment.

Leveraging this database with TPV is straightforward. TPV has been designed to allow its configuration to be split across multiple files, including files in remote locations. These are then automatically combined by TPV at runtime to create the operating configuration file. This enables the various aspects of TPV configurations to be decomposed and reused across different Galaxy deployments. To make use of the database, only its URL needs to be supplied. Once the database is linked, the resource requirements on the given Galaxy instance are always up-to-date and admins do not have to independently discover or reevaluate resourcing requirements for tools. Meanwhile, the values are constantly being adjusted by the global administrator community and anyone is welcome to contribute to the process by submitting a pull request with the desired changes.

**TPV's resource clamping**

As noted, the shared database was initially populated based on allocations that have proven to work well in the usegalaxy.* federation. These allocations however, may be too small or too great for other Galaxy installations, depending on local capacity and job sizes. To accommodate changes to specific values while still allowing admins to leverage the database, TPV has a set of override configuration options so the values can be augmented as desired. Multiple configurations were introduced to allow this augmentation at various levels of scope. Namely, values for individual tools can be overridden. However, this can be a tedious process and oftentimes administrators want a method for making sure that a job does not request resources beyond the capacity of the cluster. In such cases the job will either fail at runtime due to lack of resources, or execute successfully, albeit with degraded performance. TPV is able to handle both cases globally through support for resource limits and clamping on a per Job and per Destination basis.

Each Destination in TPV can define the maximum size of the job it is willing to accept. These are defined using `min_accepted_{cores|mem|gpus}` and `max_accepted_{cores|mem|gpu}` parameters. By defining these values, TPV will prevent tools from scheduling on destinations that do not have suitable capacity. Similarly, to accommodate the case where the default values in the shared database are requesting capacity that is in excess of what is available in the local cluster, TPV supports resource clamping. With resource clamping, an admin is able to set `max_{cores|mem|gpus}` on a Destination to designate that any value exceeding the provided ones should be clamped down. For example, if a Tool requests 64 CPUs but a Destination sets `max_cpus` to 32, the Job request will be scaled down to 32 CPUs. Depending on the tool and job properties, such jobs may not always complete successfully due to lack of resources. In combination with the `max_accepted_{cores|mem|gpu}` parameter, this risk can be reduced.

## Adoption results

Since its inception, TPV has been gradually rolled out across the three big usegalaxy.* services, starting with Galaxy Australia. Galaxy AU is the only service to date that has utilized TPV's meta-scheduling capability while the EU and US sites are only utilizing it as a source of default resource values, with more elaborate deployments planned. In this section, we report on the quantitative and qualitative experience from these deployments.

### Quantitative results

The first quantitative validation for TPV is its ability to process a large volume of jobs. Since its deployment on Galaxy AU in December 2021, over the period of the subsequent 2 years, TPV has processed over 4M jobs. It also processed about 5M jobs on the EU server (since May 2023) and about 100,000 jobs on the US server (over a two-month period starting in September 2023). For the deployments at EU and US, the time-to-schedule a job is a constant value as TPV simply performs a resource value lookup. At AU, TPV aggregates live resource status data from all available resources and invokes a rank function (more below). The average time-to-schedule on AU is measured in milliseconds, demonstrating the efficiency of its tag-based matching process.

The second quantitative validation we report is on TPV's meta-scheduling ability. Meta-scheduling is enabled by integrating TPV's rank function with live status data across all systems to which it can direct jobs. Galaxy AU is configured to operate across 6 heterogeneous, geographically distributed sites (Fig. 11). Each site has a Galaxy-specific endpoint called Pulsar that is configured to accept jobs from a central Galaxy server. Each Pulsar site represents a cluster shared with other, non-Galaxy workloads and hence the system load fluctuates independently. At Galaxy AU, the main goal for TPV's meta-scheduling rank function is to minimize job queue wait time. This has a threefold effect: (1) improve the user experience by shortening job turnaround time, and (2) increase utilization of the available allocation by directing jobs to least-loaded resources for quicker scheduling. To accomplish this, we implemented a rank function that considers the tool's destination preference (as defined in the shared resource requests database) and then orders the available resources by the CPU load on the systems (the implementation is showcased in Fig. 9). The function hence respects a tool's preference for a specific type of resource, such as big memory, and then utilizes the least loaded system.

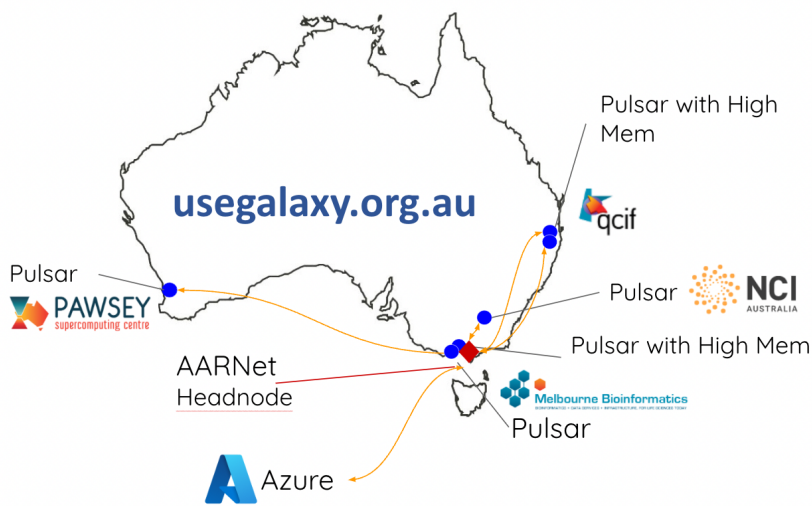

Figure 11. Sites of the resources behind Galaxy Australia (usegalaxy.org.au), physically distributed across Australia.

To observe the impact of this function, we analyzed job metrics for one year preceding TPV's deployment and one year following its deployment. During this period the number of jobs grew by 27% – from 1.4M year prior to nearly 1.8M year following. Meanwhile, the average queue wait time dropped by 15% – from 127 seconds on average to 108 seconds. This is indicative of TPV's favorable dispatch process, yielding an improved user experience. The average job runtime during the given period was about 400 seconds. A 19 second reduction in the average job queue time corresponds to a 5% shorter job processing time for an average job. Although seemingly relatively small, this is a noticeable impact at scale. From the perspective of improving utilization for the overall resource allocation, the 19 seconds reduction of job queue wait time over the 1.8M jobs that Galaxy AU processed in the year following TPV deployment, saved nearly 10,000 allocation hours. This corresponds to processing about 90,000 average user jobs instead of operating a Pulsar VM to simply monitor queued jobs.

**Qualitative results**

Qualitative results from adopting TPV focus on the system administrator managing a Galaxy server. Foremost, administrators are able to integrate TPV with their Galaxy installation by enabling it in the job configuration and then pointing its configuration to the shared resource requirement database. This database is available as a single public, online file [11]. Once enabled, the default resource requirement values for all the included tools are immediately available. In addition to the resource requirements, the database contains tool parameters, environment variables, tags, and scaling rules that are set and vetted by a community of administrators, yielding a significant reduction in general repetitiveness. Without TPV an administrator would need to configure these values for each tool manually. Given the number and diversity of tools typically installed in Galaxy, this is a time-consuming effort without a well-defined process. The shared database being a crowd-sourced effort, it is also regularly updated so a given Galaxy server is always using the most up-to-date tool values.

The second Quality-of-Life improvement that TPV brings is the ability to metaschedule jobs across multiple heterogeneous clusters, which eliminates the likelihood of jobs accumulating at any single destination as the system automatically corrects load distribution over time. Prior to TPV, this was an issue that required constant manual intervention from administrators.

A third qualitative benefit of TPV is that tag-based matching of destinations greatly simplifies administrative control over routing jobs to destinations. Examples include taking nodes offline for maintenance with a simple offline tag, or expressing preference or aversion to specific destinations.

## Summary and future perspectives

In this paper we presented Total Perspective Vortex - a library for right-sizing and meta-scheduling jobs in heterogeneous compute environments. We also presented a shared resource requirements database for a large number of popular bioinformatics tools. The captured results demonstrate that TPV is scalable and yields an efficient resource allocation. The library and the database are open source and can be readily leveraged.

Currently, TPV is available exclusively for the Galaxy application. This decision was driven by the need to address the challenge of improving utilization of the available resources across the usegalaxy.* federation in light of its growing adoption. Today, TPV can be configured for use by any Galaxy application via a set of straightforward configuration files. However, the job scheduling results TPV yields are applicable beyond the Galaxy ecosystem and even beyond bioinformatics. Hence one of the future goals is to make TPV more generally applicable with one potential avenue being to develop a general-purpose API and deploy it as a service that other workflow management systems can adopt.

The second major contribution of this paper - the resource requirements database is likewise applicable beyond the Galaxy ecosystem. However, maintaining the resourcing requirements for a growing number of tools is challenging even for a global community of administrators. A key advancement for scaling the scope of the database would be to automate the process of defining and updating resource requirements. The usegalaxy.* federation has collected detailed runtime data on millions of jobs. In the past, attempts have been made to use machine learning to predict resource requirements based on this historical data captured in Galaxy [12], but these efforts have been challenged by the noisy nature of the data. Therefore, rather than adopting an all-encompassing machine-learning approach, we think there is space to follow a more incremental approach, such as automated bracketing of input sizes for tools that scale linearly with input, and semi-automate the lifecycle of this database.